\documentclass[useAMS,usenatbib,usegraphicx]{mn2e}

\title[Rocky Planetesimals Orbiting Two Hyads]{Evidence of Rocky Planetesimals Orbiting Two Hyades Stars}

\author[J. Farihi et al.]{J. Farihi$^1$\thanks{E-mail: jfarihi@ast.cam.ac.uk}\thanks{STFC Ernest Rutherford Fellow}, B. T. G\"ansicke$^2$,
D. Koester$^3$\\
$^1$Institute of Astronomy, University of Cambridge, Cambridge CB3 0HA\\
$^2$Department of Physics, University of Warwick, Coventry CV4 7AL\\
$^3$Institut f\"ur Theoretische Physik und Astrophysik, University of Kiel, 24098 Kiel, Germany}

\begin{document}

\date{}

\maketitle

\label{firstpage}

\begin{abstract}

The Hyades is the nearest open cluster, relatively young and containing numerous A-type stars; its known age, distance, and 
metallicity make it an ideal site to study planetary systems around $2-3\,M_{\odot}$ stars at an epoch similar to the late heavy 
bombardment.  {\em Hubble Space Telescope} far-ultraviolet spectroscopy strongly suggests ongoing, external metal pollution 
in two remnant Hyads.  For ongoing accretion in both stars, the polluting material has $\log[n({\rm Si})/n({\rm C})]>0.2$, is more 
carbon deficient than chondritic meteorites, and is thus rocky.  These data are consistent with a picture where rocky planetesimals 
and small planets have formed in the Hyades around two main-sequence A-type stars, whose white dwarf descendants bear the 
scars.  These detections via metal pollution are shown to be equivalent to infrared excesses of $L_{\rm IR}/L_*\sim10^{-6}$ in the 
terrestrial zone of the stars.

\end{abstract}


\section{INTRODUCTION}

Despite the proliferation of confirmed exoplanets, and continuing successes in novel methods and instruments designed for their 
detection, planetary systems in star clusters remain elusive.  Of the roughly 800 known exoplanets, only {\em four} are known  to orbit 
stars in clusters.  The first two exoplanets identified in clusters were giant planets orbiting giant stars \citep{sat07,lov07}; one of those 
planets ($\epsilon$\,Tau\,b) was discovered in the Hyades.  This relatively tiny number has only recently doubled with the radial velocity
detection of two hot Jupiters around Sun-like stars in Praesepe \citep{qui12}.  

This situation is unfortunate because star clusters can, in principle, provide a sound statistical basis for planet population studies.  A 
cluster offers uniform distance, age, and metallicity, and typically has a well-studied mass function and local environment \citep{coc02}.  
Specifically, clusters allow studies of planet formation as a function of stellar mass, as this is one of the few independent variables, and 
arguably the most significant.  With its super-solar metallicity, the Hyades is an excellent hunting ground for giant planets, but a precision 
radial velocity survey of nearly 100 solar- and low-mass main-sequence stars resulted in no detections \citep{pau04}.  A similar search 
of about 90 dwarf, subgiant, and giant stars in M67 also produced no planet candidates \citep{pas12}.  It may be that the lack of planet 
detections (relative to field stars) is a telltale, and that clusters have an impact on both the planet formation process or their longer-term 
survival, but a lack of sensitivity due to youthful stellar activity is a distinct possibility for the null results of these radial velocity searches.

Debris disks around cluster stars provides complimentary information, indicating the presence of planetesimal populations that are 
typically Kuiper Belt analogs \citep{wya03,zuc01,hol98}, and sometimes containing the dynamical signatures of planets \citep{kal05}.
Several examples of dusty, intermediate- and solar-mass, main-sequence stars have been identified in the Hyades and similarly 
young open clusters \citep{urb12,cie08}; their (infrared) spectral energy distributions are comparable to those observed toward field 
stars with circumstellar dust \citep{su06,bry06}, and are thus compatible with outer system, icy planetesimal belts.  To date, only the
unusually strong and warm infrared excess detected around a solar-mass Pleiad \citep{rhe08} has furnished evidence of debris 
within the inner planetary system of a cluster star, and supports a picture of a dynamically active, terrestrial planetary system.

\begin{figure}
\includegraphics[width=86mm]{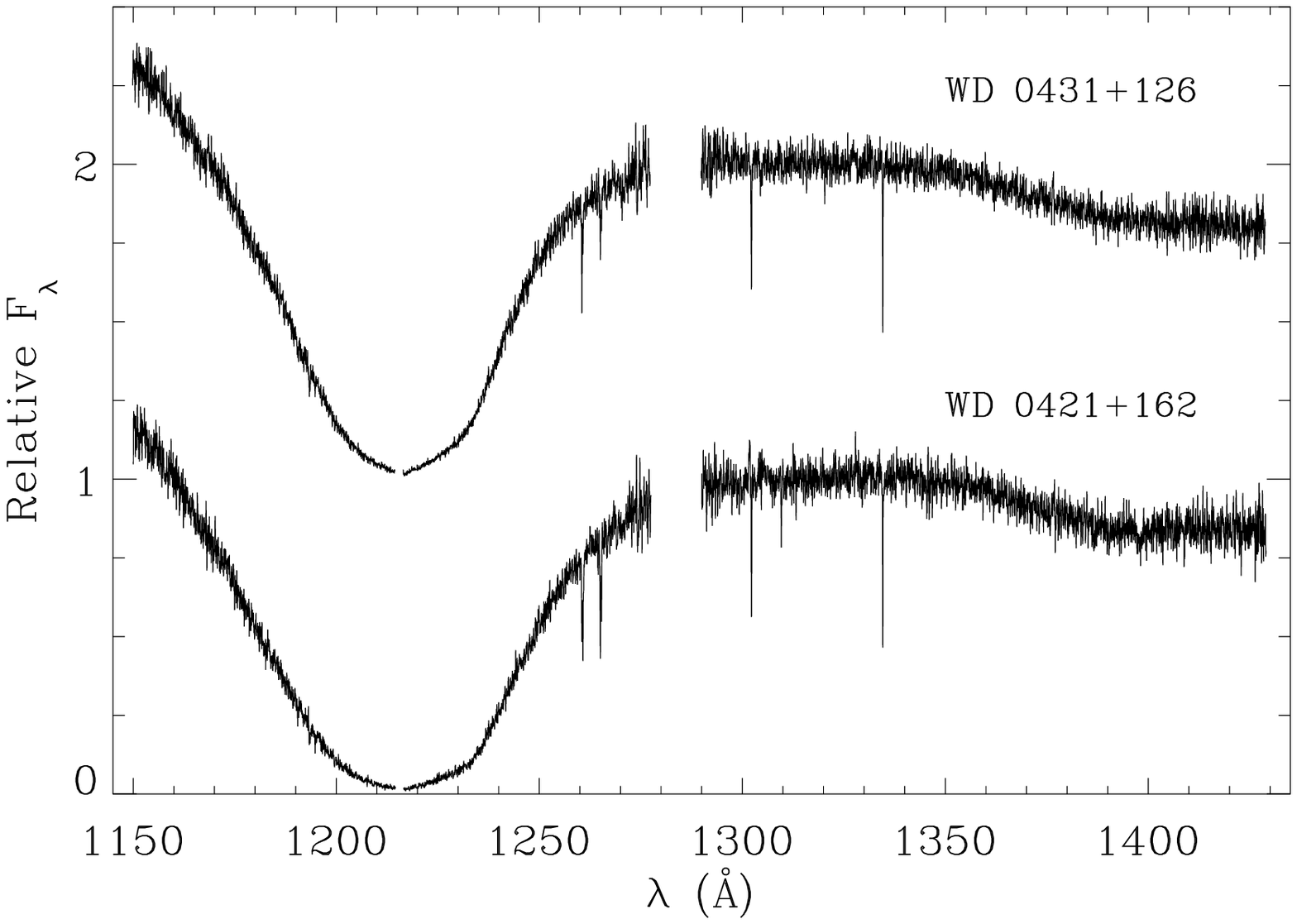}
\caption{COS spectra of the two white dwarf Hyads.  The data have been smoothed by a 5-pixel Gaussian to a resolution near 
0.05\,\AA \ (the break in wavelength coverage is due to the gap between detectors).  See Figure \ref{fig2} for feature details and
model fits.
\label{fig1}}
\end{figure}

This paper reports the detection of silicon-rich material that is consistent with rocky planetary debris around two descendants of 
intermediate mass (late B- or A-type) stars in the Hyades open cluster.  The discovery was made via far-ultraviolet spectroscopy 
of two white dwarf Hyads: WD\,0421$+$162 (EG\,36) and WD\,0431$+$126 (EG\,39).  The spectra of both stars contain absorption 
lines of Si\,{\sc ii}, and upper limits for volatile elements such as carbon, that suggest ongoing accretion from a young but evolved 
terrestrial planetary system.  The observations and data analysis are described in \S2, and properties of the planetary debris are 
discussed in \S3, where it is also shown that white dwarf metal pollution is more sensitive than other methods of exo-terrestrial 
debris detection.  Whereas this paper reports the ultraviolet observations of two white dwarf Hyads, Zuckerman et al. (2013, 
submitted to ApJ) report an optical study of atmospheric metals in Hyades white dwarfs.

\section{OBSERVATIONS AND ANALYSIS}

\subsection{Ultraviolet Spectra}

The two Hyads were selected as part of {\em Hubble Space Telescope (HST)} Cycle 18 Snapshot program 12169 to search 
for external metal pollution in 17\,000\,K $< T_{\rm eff} < 25\,000$\,K hydrogen-dominated (DA-type) white dwarfs; their cluster 
membership was incidental.  Both stars were observed for 400\,s with the Cosmic Origins Spectrograph (COS) using the G130M 
grating and a central wavelength setting at 1291\,\AA, covering $1130-1435$\,\AA.  The data were processed and calibrated with 
{\sc calcos} 2.15.6, and are shown in Figure \ref{fig1}.

\begin{figure*}
\includegraphics[width=172mm]{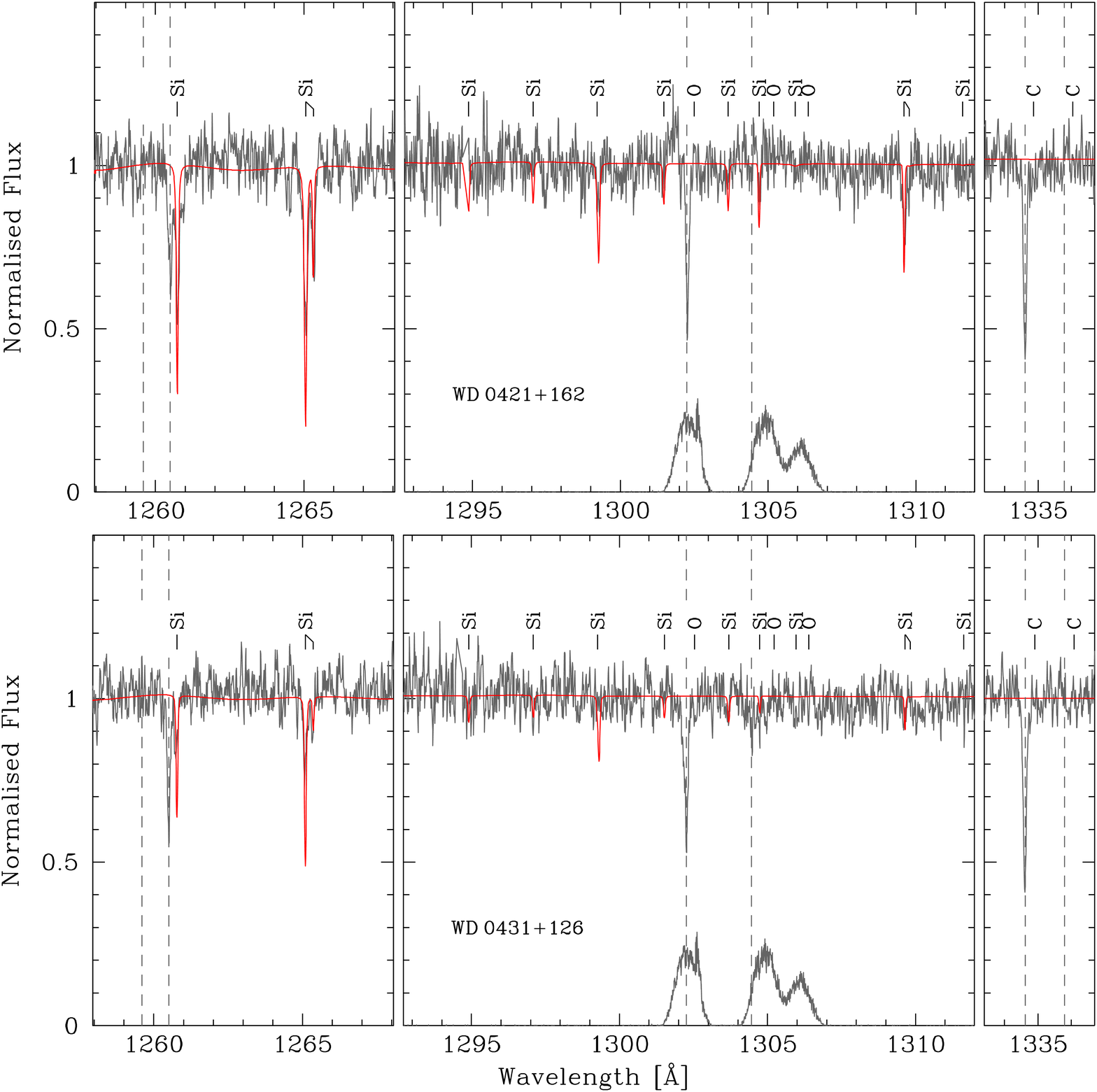}
\caption{The strongest features detected in the normalized COS spectra (gray) of the two Hyads, together with the best fitting model
spectra (red).  Photospheric lines of Si\,{\sc ii} are seen in both stars at vacuum wavelengths 1190.4, 1193.3, 1194.5, 1260.4, 1264.7, 
1265.0\,\AA, while 0421$+$162 also exhibits 1309.3, 1309.5\,\AA \ lines of the same ion.  Interstellar resonance lines are present at 
1260.4\,\AA \ (Si\,{\sc ii}, but blue-shifted in both stars by 60\,km\,s$^{-1}$ with respect to the corresponding photospheric lines), 
1302.2\,\AA \ (O\,{\sc i}), and 1334.5\,\AA \ (C\,{\sc ii}), and are indicated by dashed lines.  Similarly indicated are additional interstellar 
lines at 1259.5\,\AA \ (S\,{\sc ii}) and 1335.7\,\AA \ (C\,{\sc ii}) often seen in far-ultraviolet spectra of nearby white dwarfs; their absence 
here illustrates the low interstellar column density along the line of sight to the cluster.  Geocoronal airglow of O\,{\sc i} at 1302.2, 1304.9,
and 1306.0\,\AA \ can contaminate COS spectra to some degree, and typical airglow line profiles are shown in the middle panel scaled 
to an arbitrary flux.
\label{fig2}}
\end{figure*}

The spectra reveal the Stark-broadened Ly$\alpha$ profile intrinsic to DA stars, plus several photospheric lines of Si\,{\sc ii}.  Also 
visible in each of the spectra are three relatively strong interstellar (resonance) lines, one each of C\,{\sc ii}, O\,{\sc i}, and Si\,{\sc ii}.  
In both stars, the latter interstellar line is accompanied by a relatively strong photospheric line, which is significantly red-shifted ($
\Delta v\approx60$\,km\,s$^{-1}$) as to be fully separated from the interstellar component at the instrumental resolution (Figure 
\ref{fig2}).  Raw signal-to-noise was calculated in the relatively flat region between 1310 and 1330\,\AA, yielding 11.4 for 0421$
+$162 and 12.7 for 0431$+$126.

\subsection{Atmospheric Parameters and Metal Abundances}

Using available $UBVJ$ photometry and the cluster center parallax \citep{per98}, spectral models were fitted to the measured fluxes
and stellar parameters were derived independent of spectroscopy.  The effective temperatures and surface gravities calculated in this
way are listed in Table \ref{tbl1} and the resulting masses are listed as $M_{\rm cl}$ in Table \ref{tbl2}.  The most significant source of 
uncertainty in this method is the parallax, which applies to the cluster core; while the two white dwarfs appear to be well within the
central region based on their distance from the core in the plane of the sky, their actual radial offset is relatively uncertain.

The COS spectra were analyzed as detailed in \citet{gan12}, using the input physics of \citet{koe10} but with $\log\,g$ fixed to the values 
from the photometry and cluster parallax.  Atmospheric parameters determined for both stars from the ultraviolet data are listed in Table
\ref{tbl1}, where the reported errors are statistical only.  Actual errors can be estimated by comparison with previous determinations that 
use similar or identical models, but are based on optical spectra.  There exist several spectroscopic analyses for both 0421$+$162 and 
0431$+$126, the most recent of which are based on single-order optical spectra \citep{gia11}, and optical echelle data \citep{koe09}; 
both are listed in Table \ref{tbl1}.  Overall, there is good agreement between the various methods.

Importantly, it should be emphasized that {\em the exact choice of $T_{\rm eff}$ and $\log\,g$ does not significantly alter the derived 
metal abundances}.  The uncertainties give rise to abundance errors smaller than the typical measurement error (0.1\,dex), and the 
metal-to-metal ratios are essentially unaffected \citep{gan12}.  It is still worthwhile, however, to consider the most accurate white dwarf
parameters, as these can be linked to their main-sequence progenitors, and thus provide insight on the rocky planetary systems they 
construct.  The adopted $T_{\rm eff}$, $\log\,g$, $M_{\rm wd}$ in Table \ref{tbl2} are the unweighted average of all the independent 
values listed in Table \ref{tbl1}, while $M_{\rm sp}$ is the mass derived from the unweighted average of the two optical spectroscopic
datasets.  Table \ref{tbl2} also lists main-sequence progenitor masses for the two Hyads, derived using appropriate initial-to-final mass 
relations \citep{wil09,kal08}.

The metal absorption features in the COS spectra were analyzed using the $T_{\rm eff}$ and $\log\,g$ listed at the top of Table \ref{tbl2}.  
Silicon abundance ratios, relative to hydrogen, were calculated for both stars, as well as upper limits for carbon based on several strong 
lines, including the C\,{\sc ii} resonance line.  From the derived abundances and upper limits, diffusion fluxes were calculated following 
\citet{gan12}; these are exactly the accretion rates and limits for each element and are listed in Table \ref{tbl1}.  This is the correct approach 
for these warm DA stars, which essentially lack convection zones, and diffusion timescales change as a function of optical depth.  In these 
white dwarfs, the timescale for metals to sink is very short (order of days) and a steady state assumption is safe (for a detailed discussion, 
see \citealt{gan12}).

\begin{table}
\begin{center}
\caption{Atmospheric Parameters from Photometry and Cluster Parallax, Optical and Ultraviolet Spectroscopy\label{tbl1}} 
\begin{tabular}{@{}cccc@{}}
\hline

WD				&$T_{\rm eff}$				&$\log\,g$					&Method\\
				&(K)						&(cm\,s$^{-2}$)				&\\

\hline

0421$+$162		&18\,676($\phantom{0}$130)	&8.05(02)					&1\\
				&19\,364($\phantom{00}$40)	&8.08(01)					&2\\
				&20\,010($\phantom{0}$315)	&8.13(05)					&3\\
				&18\,918($\phantom{000}$8)	&8.05$^{\phantom{0000}}$	&4\smallskip\\
				
0431$+$126		&21\,000(1022)			&8.14(06)					&1\\
				&20\,929($\phantom{00}$40)	&8.09(01)					&2\\
				&21\,890($\phantom{0}$346)	&8.11(05)					&3\\
				&20\,992($\phantom{000}$9)	&8.14$^{\phantom{0000}}$	&4\\

\hline
\end{tabular}
\end{center}

Method: (1) $UBVJ$ photometry combined with cluster parallax; (2) Optical echelle spectroscopy \citep{koe09} with updated
models (S2.1); (3) Single-order optical spectroscopy \citep{gia11}; (4) COS spectroscopy with $\log\,g$ fixed.
\smallskip

\end{table}

\subsection{Photospheric Velocities and Gravitational Redshifts}

Both stars were also observed in the SPY survey \citep{nap03}, on two separate nights each with UVES, and these spectra were 
downloaded from the VLT archive and analyzed to search for additional metals lines and to calculate radial velocities.  There is no 
evidence of Ca\,{\sc ii} K 3933.7\,\AA \ or Mg\,{\sc ii} 4481.2\,\AA \ absorption, which are the strongest metal transitions in the optical 
for white dwarfs in this temperature range.  The line velocities of H$\alpha$ and H$\beta$ (the sum $v_{\rm rad} + v_{\rm gr}$) were 
measured for both stars, in each of their datasets and are listed in Table \ref{tbl2}.  The measurements for H$\beta$ agree within 2\% 
of that found for H$\alpha$, but the NLTE core of the latter line makes it more reliable and thus only those velocities were used.

Notably, the average, heliocentric-corrected velocities from fits to three Si\,{\sc ii} lines in the COS spectra of both stars, listed in Table 
\ref{tbl2}, agree remarkably well with the Balmer line velocities, and demonstrate unambiguously that the metal lines are photospheric.  
The velocities derived from the interstellar C\,{\sc ii} and Si\,{\sc ii} lines are consistent with those measured for the local interstellar cloud 
toward several Hyades stars \citep{red01}.

If both white dwarfs are moving at a radial velocity similar to the cluster center, then their remaining velocity components can be used 
to establish their masses and radii via gravitational redshifts plus a (theoretical) mass-radius relation \citep{koe87}.  This is the only 
available method for deriving single white dwarf masses that does not rely on atmospheric models.  The cluster core radial velocity is 
$+38.6$\,km\,s$^{-1}$, the two stars are 0.9 and 3.4\,pc distant from the cluster center in the plane of the sky, and are therefore likely to 
be within the $r<10$\,pc defined cluster center \citep{per98}.  Gravitational redshifts were derived from the UVES data and white dwarf
masses were calculated adopting the cooling models of \citet{hol06} and the mass-radius relations of \citet{pan00}.  Table \ref{tbl2}
summarizes the results and yields very similar masses for the two stars, listed as $M_{\rm gr}$; these measurements compare favorably 
to those determined in previous Hyades studies \citep{cla01,rei96,weg89}.

\begin{table}
\begin{center}
\caption{Results Summary\label{tbl2}} 
\begin{tabular}{@{}lrr@{}}
\hline

WD									&0421$+$162			&0431$+$126\\

\hline

Adopted Parameters:					&&\\
$T_{\rm eff}$ (K)						&19242(586)			&21202(459)\\
$\log\,[g \ ({\rm cm\,s^{-2}}$)]				&8.09(04)				&8.11(03)\medskip\\

Velocities (km\,s$^{-1}$):					&&\\

$v_{{\rm H}\alpha,1}$					&74.8(0.6)			&73.7(0.6)\\
$v_{{\rm H}\beta,1}$						&74.1(1.3)			&72.6(1.3)\\
$v_{{\rm H}\alpha,2}$					&76.3(0.6)			&76.6(0.6)\\
$v_{{\rm H}\beta,2}$						&78.1(1.3)			&77.2(1.3)\\
$v_{\rm Si}$$^\dag$						&77.1				&79.0\\
$v_{gr}$								&37.0(0.6)			&36.6(0.6)\medskip\\

Stellar Masses  ($M_{\odot}$):				&&\\

$M_{\rm cl}$							&0.65(01)				&0.71(04)\\
$M_{\rm sp}$							&0.69(02)				&0.69(02)\\
$M_{\rm gr}$ 							&0.70(01)				&0.70(01)\\
$M_{\rm wd}$							&0.67(02)				&0.69(02)\\
$M_{\rm ms}$ 							&2.5(0.2)				&2.7(0.2)\medskip\\

Debris Properties:						&&\\

$\log[n({\rm Si})/n({\rm H})]$				&$-7.5$				&$-8.0$\\	
$\log[n({\rm Si})/n({\rm C})]$				&$>0.7$				&$>0.2$\\	
$\log[\dot M_{\rm Si}$ (g\,s$^{-1}$)]			&5.3					&4.8\\
$\log[\dot M_{\rm Z}$ (g\,s$^{-1}$)]$^\ddag$	&6.2					&5.6\\

\hline
\end{tabular}
\end{center}

{\em Note}.  $M_{\rm cl}$ is the mass derived assuming the cluster parallax, $M_{\rm sp}$ is the mass derived via spectroscopy, 
$M_{\rm gr}$ is the mass derived by gravitational redshift, $M_{\rm wd}$ is the adopted white dwarf mass, and $M_{\rm ms}$ is 
the main-sequence progenitor mass.  Gravitational redshifts are derived by assuming $v_{\rm rad}=38.6$\,km\,s$^{-1}$ (See \S2.2 
and 2.3 for details).
\smallskip

$^\dag$Errors are a few km\,s$^{-1}$.
\smallskip

$^\ddag$Calculated assuming silicon represents 0.16 of the total mass, as in the bulk Earth (\S3).

\end{table}

\section{RESULTS AND DISCUSSION}

\subsection{Not Interstellar Matter}

Before discussing the implications of detecting photospheric silicon in the COS spectra of these remnant Hyads, it is appropriate to
review the arguments against an interstellar origin for these lines.  First, while absorption from the local interstellar cloud is detected 
via the strong 1260.4\,\AA \ resonance line, the additional lines at 1264.7 and 1265.0\,\AA \ arise from an excited state 0.035\,eV above 
the ground state, which is unpopulated in the interstellar medium, and hence must be photospheric.  Second, if the white dwarfs have 
accreted interstellar material, one would expect to detect photospheric carbon at Si/C broadly consistent with the solar value.  As 
shown below, the non-detection of carbon implies that infalling debris is at least an order of magnitude higher in Si/C than expected 
for the interstellar medium.  Third, even for the relatively low accretion rate of $\dot M_{\rm Si}=10^{4.8}$\,g\,s$^{-1}$ in 0431$+$126, 
the infall of interstellar gas would be dominated by {\em hydrogen} at $\dot M_{\rm H}\geq10^{7.8}$\,g\,s$^{-1}$.  For Eddington type
(gravitational) accretion at relative speeds of 40\,km\,s$^{-1}$ this would require $\rho_{\rm H}\sim100$\,cm$^{-3}$ \citep{far10a}.

\subsection{Can Silicon be Supported by Radiative Levitation?}

For DA white dwarfs of $T_{\rm eff}\approx20\,000$\,K, the detection of atmospheric metals has previously been an unambiguous 
sign of ongoing accretion.  Only high levels of metal pollution are detectable with optical spectroscopy of warm DA stars, via calcium 
or magnesium absorption lines, and the corresponding heavy element masses cannot be radiatively sustained in any atmospheric 
models \citep{koe06}.  The sensitive ultraviolet data obtained with COS reveal relatively modest levels of atmospheric silicon, an 
element that could potentially be susceptible to radiative levitation in this range of effective temperatures, unlike calcium and 
magnesium.

The amount of atmospheric silicon that can be supported by stellar radiation is not known, and previous model comparisons with
metals in $T_{\rm eff}>25\,000$\,K white dwarfs have not yet been quantitatively successful \citep{cha95b}.  The only available model
at lower temperatures predicts that surface silicon abundances of [Si/H] $=-8.0$ are maintained by radiative acceleration at 20\,000\,K 
and $\log\,g=8.0$ \citep{cha10}; 0421$+$162 and 0431$+$126 have [Si/H] $=-7.5$ and $-8.0$ respectively.

However, there are at least three problems with the picture in which silicon is radiatively supported in both of these white dwarfs.  First, 
the {\em cooler} of the two stars has {\em more} atmospheric silicon, in contrast to model predictions in which levitation efficiency scales
with luminosity.  Second, in the full COS Snapshot sample, there are numerous stars in the same temperature range with detected silicon 
abundances lower than the two Hyads (G\"ansicke et al. 2013, in preparation; \citealt{koe12}).  Specifically, there are at least 11 stars with 
measured [Si/H] $<-8.0$, and seven of these are predicted to have higher levitation efficiency than both 0421$+$162 and 0431$+$126
\citep{koe12}.  Third, silicon cannot have been supported in the atmosphere of these two white dwarfs for their entire cooling age; near 
70\,000\,K any and all primordial silicon will sink \citep{cha95a}, and thus the silicon must be external.  Radiative levitation of silicon is
therefore not supported by the data at the observed atmospheric parameters and abundances.

\begin{figure}
\includegraphics[width=86mm]{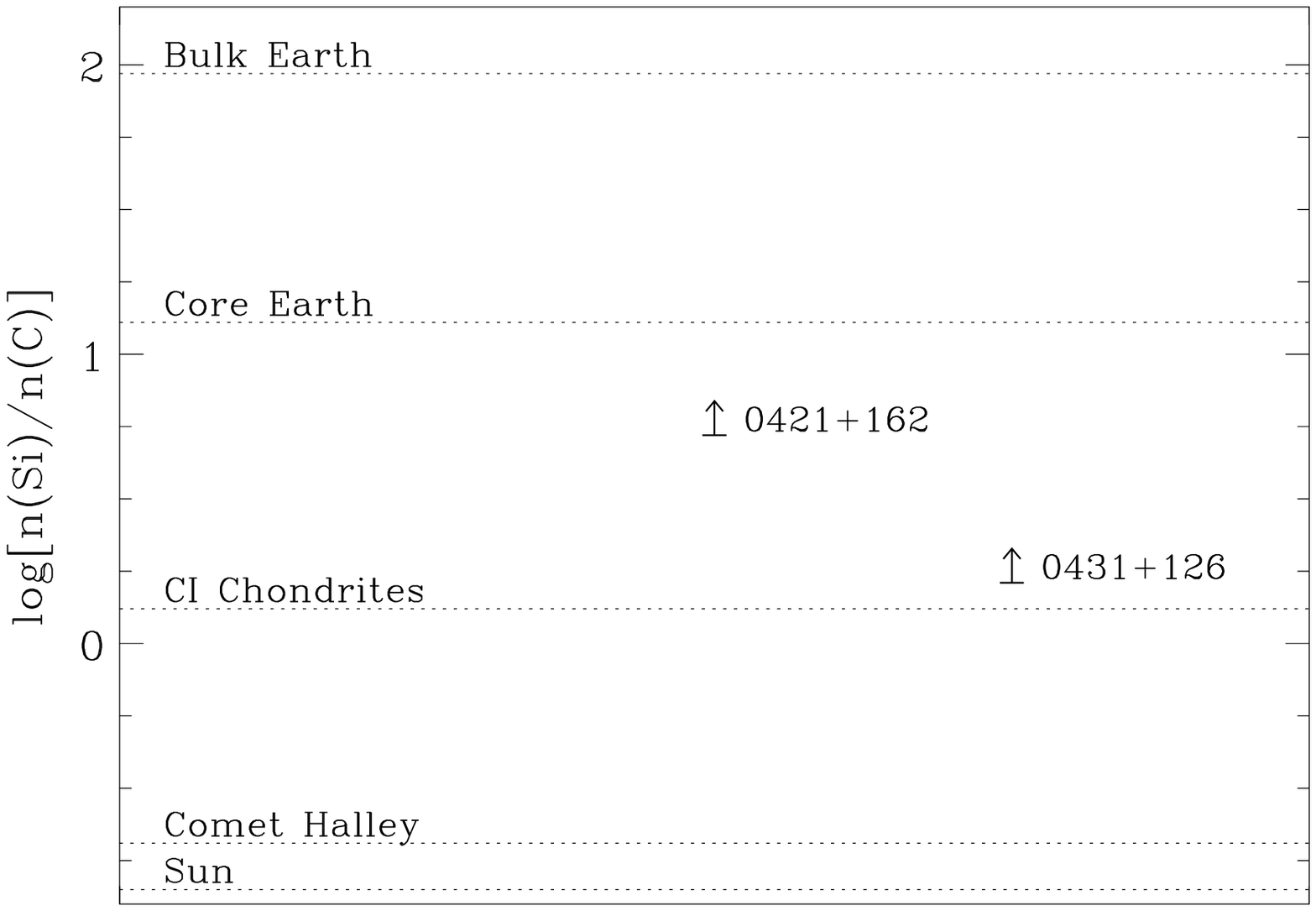}
\caption{Lower limits to the Si/C abundances in the material polluting the two Hyades white dwarfs, assuming accretion is ongoing.
Also shown are the same ratios for the Sun, Halley, chondrites, the core and bulk Earth \citep{lod03,mcd00,lod98}.  These lower limits
indicate that currently infalling material in both stars must be rocky; the Si/C ratios are more volatile-depleted than those in chondritic 
meteorites, suggesting formation in the terrestrial zone of their respective planetary systems.
\label{fig3}}
\end{figure}

\subsection{Super Chondritic Si/C}

Figure \ref{fig3} plots the Si/C number ratio for each of the stars, and compares it to several Solar System benchmarks.  The key result
is that material being currently accreted must be more carbon-deficient than chondritic meteorites and thus rocky.  In contrast, planetary 
debris that originates in an extrasolar Kuiper Belt analog is expected to have a high fraction of volatiles like carbon, as observed in comets such as Halley \citep{mum93}; during the giant phases of the progenitor star, volatile elements should not be significantly depleted within 
$r\ga50$\,km planetesimals orbiting beyond 30\,AU \citep{jur10}.  Thus, planetary material currently being accreted by these two 
descendants of A-type stars in the Hyades was likely formed in their inner regions, and is thus analogous to asteroidal material, i.e.\ 
exo-terrestrial.

The total heavy element accretion rates for the stars, $\dot M_{\rm Z}$ in Table \ref{tbl2}, are calculated assuming silicon is 0.16 by 
mass as in the bulk Earth \citep{mcd00,mcd95,all95}, as this has been shown to most closely reflect the total composition where eight 
or more metals are detected \citep{zuc10}.  These inferred total accretion rates are between 10 and 100 times lower than found for any
other DAZ star of similar effective temperature, and this is due to the superior sensitivity of COS ultraviolet data (hence the motivation 
for the COS Snapshot).  

Assuming these stars are accreting calcium and magnesium in bulk Earth proportions relative to silicon, the photospheric abundances 
would be [Ca/H] $=-8.7,-9.3$ and [Mg/H] $=-7.3,-7.9$ for 0421$+$162, 0431$+$126 respectively.  Model spectra calculated with these 
abundances have Ca\.{\sc ii} 3934\,\AA \ and Mg\,{\sc ii} 4482\,\AA \ line widths around 1\,m\AA \ and would therefore not be detectable 
with current ground-based facilities.  While such detections should be possible with ELTs, the Mg\,{\sc ii} resonance lines in the 
near-ultraviolet are predicted to be significantly larger, at $50-100$\,m\AA.

\subsection{Rocky Debris Detection: Pollution vs. Infrared Excess}

{\em Spitzer} studies of metal-enriched white dwarfs have shown that only those circumstellar disks with infall rates above $10^{8}
$\,g\,s$^{-1}$ are capable of producing detectable infrared excesses \citep{far09,jur07}, and the silicon-bearing Hyads keep to this
trend.  All eight of the classical, single Hyades (super)cluster white dwarfs were observed with cryogenic IRAC -- to search for young
and still-warm giant planets -- but no infrared excesses were detected \citep{far08}.  

For polluted white dwarfs with infrared-detected disks, the thermal $F_{\nu}$ continuum peaks near 5\,$\mu$m from $T\approx
1000$\,K dust \citep{gir12,von07}, in stark contrast with dusty main-sequence stars such as Vega and $\beta$\,Pic.  The white
dwarf disks have relatively large fractional infrared luminosities, ranging from $0.001$ to $0.03$, including at least six stars with 
$L_{\rm IR}/L_*>1\%$ \citep{far12,far10b}.  The fractional excesses tend towards smaller values with increasing stellar effective 
temperature \citep{far11}, possibly due to a smaller available area to un-sublimated dust \citep{far12} within the finite Roche limit 
of the star, where the debris is likely generated by asteroids perturbed onto star-crossing orbits \citep{ver13,bon12,deb12,deb02}, 
and subsequently tidally disrupted \citep{jur03}.  These data can be used to estimate $L_{\rm IR}/L_*$ for the two polluted Hyads, 
by assuming disk mass scales with accretion rate, and extrapolating from infrared detections.  Taking $L_{\rm IR}/L_*\sim0.001$ for 
$10^{9}$\,g\,s$^{-1}$ \citep{far11} from the known $T_{\rm eff}\sim20\,000$\,K white dwarfs with dust, the corresponding fractional 
dust luminosities would be $10^{-5.8}$ and $10^{-6.4}$ for 0421$+$162 and 0431$+$126 respectively.

Hence, these detections of exo-terrestrial debris via atmospheric metal pollution in white dwarfs are at least as sensitive as 
equivalent infrared observations of main-sequence stars.  For A-type stars, which are the likely progenitors of the two Hyads studied 
here, $L_{\rm IR}/L_*\sim10^{-5}-10^{-6}$ sensitivities are readily achieved via 24 and 70\,$\mu$m photometry \citep{su06}, but are 
mainly insensitive to dust within 10\,AU \citep{wya08}.  Only a small fraction of main-sequence stars display infrared excesses consistent 
with warm debris in the terrestrial zones of their host stars \citep{ken12,mel10}, and by necessity they tend to have spectacular infrared 
excesses \citep{rhe08,son05}.  For observations that probe within several AU, such as 12\,$\mu$m {\em IRAS} or {\em WISE} photometry, 
dust detections  typically require $L_{\rm IR}/L_*\ga10^{-4}$.  As shown above, it is likely that white dwarfs enable detections at least two
orders of magnitude smaller than this via pollution, and are thus a highly sensitive probe of terrestrial planetary debris around nearby stars.

\section{SUMMARY}

Sensitive ultraviolet observations with {\em Hubble} COS suggest the ongoing accretion of silicate-rich and carbon-poor circumstellar
material at two white dwarf descendants of intermediate-mass (A-type) stars in the Hyades open cluster.  In each case, currently infalling
material is found to be more carbon-deficient than CI chondrites, and consistent with terrestrial-like planetesimals at 625\,Myr, which is
approximately the timescale of the late heavy bombardment in the Solar System.  This evidence supports the idea that these two Hyads 
had the rocky building blocks necessary for the formation and retention of terrestrial planets.

The study of rocky exoplanetary material via white dwarfs has great potential.  It is the only method whereby the bulk composition 
of entire planetesimals (and possibly large bodies akin to planetary embryos) can be ascertained.  Infrared observations can reveal 
material in the terrestrial zone of nearby stars, but the requirement for large flux excesses -- and the dearth of examples despite large
searches -- implies that white dwarf pollution may already have yielded more detections of material in this region.  At least 20\% to 30\%
of cool white dwarfs show evidence for refractory element pollution \citep{zuc10,zuc03}, compared to less than 1\% of main-sequence 
stars with warm debris.  In fact, {\em two of the three nearest white dwarfs are metal-polluted}: vMa\,2 and Procyon\,B \citep{far13}.  
The discovery of what are likely rocky planetary systems in the Hyades highlights the power of the archaeological approach to 
terrestrial exoplanetary science.

\section*{ACKNOWLEDGMENTS}

The authors thank S. Redfield for useful conversations regarding interstellar matter in the vicinity of the Hyades, and the anonymous
referee for comments that improved the manuscript.  Balmer lines in the atmospheric models were calculated with the modified Stark 
broadening profiles of \citet{tre09} kindly made available by the authors.  This work is based on observations made with the {\em Hubble
Space Telescope} which is operated by the Association of Universities for Research in Astronomy under NASA contract NAS 5-26555.  
These observations are associated with program 12169.   J. Farihi gratefully acknowledges the support of the STFC via an Ernest 
Rutherford Fellowship, and B. T. G\"ansicke was supported by the STFC through a rolling grant at Warwick University.

\label{lastpage}

\end{document}